\documentstyle[aps,twocolumn]{revtex}
\begin{document}
\draft
\preprint{quant-ph/0009030}
\title{One- and two-dimensional $N$-qubit systems in
capacitively coupled quantum dots}
\author{Tetsufumi Tanamoto\cite{mail}}
\address{Corporate R \& D Center,
Toshiba Corporation, 1, Komukai Toshiba-cho,
Saiwai-ku, Kawasaki 212-8582, Japan}
\date{\today}
\maketitle

\begin{abstract}
Coulomb blockade effects in capacitively coupled quantum dots 
can be utilized for constructing an $N$-qubit system with 
antiferromagnetic Ising interactions.  Starting from the tunneling 
Hamiltonian, we theoretically show that the Hamiltonian for a weakly coupled 
quantum-dot array is reduced to that for nuclear magnetic resonance 
(NMR) spectroscopy. Quantum operations are carried out  
by applying only electrical pulse sequences. Thus   
various error-correction methods developed in NMR spectroscopy 
and NMR quantum computers
are applicable without using magnetic fields.
A possible measurement scheme in an $N$-qubit system is quantitatively discussed.  
\end{abstract}


\narrowtext
Quantum computers have been widely investigated 
from many perspectives
\cite{DiVincenzo,Bennett,Lloyd,Kane,Loss,DiVincenzo2,Averin,Shnirman,Mooij,Barenco,Openov,Tanamoto}. 
A quantum-dot array is a promising candidate 
for the basic element of a quantum computer from the viewpoint 
of technological feasibility\cite{Zhitenev,Oosterkamp}. 
It is noteworthy that controlling of lateral and vertical order in 
self-organized quantum-dot superlattices has been realized\cite{Springholz}.
Although a spin-based quantum dot computer has been intensively discussed\cite{Loss,DiVincenzo2}, 
it seems that the quantum dot system using charged states is more accessible 
because the latter will be able to be constructed of various materials other 
than III-V group materials such as GaAs.
From this perspective, several authors investigated the quantum computer based on 
the charged states\cite{Barenco,Openov,Tanamoto}. 
In Ref.\cite{Tanamoto}, we discussed the quantum dot computer 
in the limit of a free-electron approximation. The free-electron approximation 
will be valid only when the interdot tunneling is strong and Coulomb blockade 
is suppressed, or the size of the quantum dots is as small as that of artificial atoms.  
Generally speaking, however, given the current state of technology,
it seems that it will not be possible to make quantum dots as small 
as atomic order in the near future. 
Moreover, controlled-NOT (CNOT) operation 
for a quantum computer based 
on charged states 
has been discussed only in the two neighboring qubits 
so far; 
a scheme for constructing more than two qubits on a device  
remains unclear. 

In this paper we advance the analysis of the quantum computer based on charged states 
and show a general $N$-qubit scheme.
Starting from tunneling Hamiltonian in the Coulomb blockade regime, 
we demonstrate that the Hamiltonians for one- and two-dimensional arrays  
of weakly coupled quantum dots are reduced to those for  
standard nuclear magnetic resonance (NMR) spectroscopy\cite{Ernst}. 
This enables any quantum computation to be described by electric pulse sequences. 
We assume that interdot tunneling is not so strong that the capacitance between two 
quantum dots is defined and we can use the Hamiltonian which allows the 
electrostatic energy of the quantum-dot structure to be described in 
terms of the capacitances of individual dots and gate voltages. 
This type of quantum-dot system was discussed experimentally 
by Livermore et al\cite{Livermore}. 
The interaction between qubits is derived from the Coulomb 
interaction between the excess electrons in the quantum dots. 
We also numerically illustrate the measurement process using  the field-effect, 
which is considered to be 
a sensitive measurement method for electronic states\cite{Landauer1}. 
We set $e$=1 and $k_{\rm B}=1$.

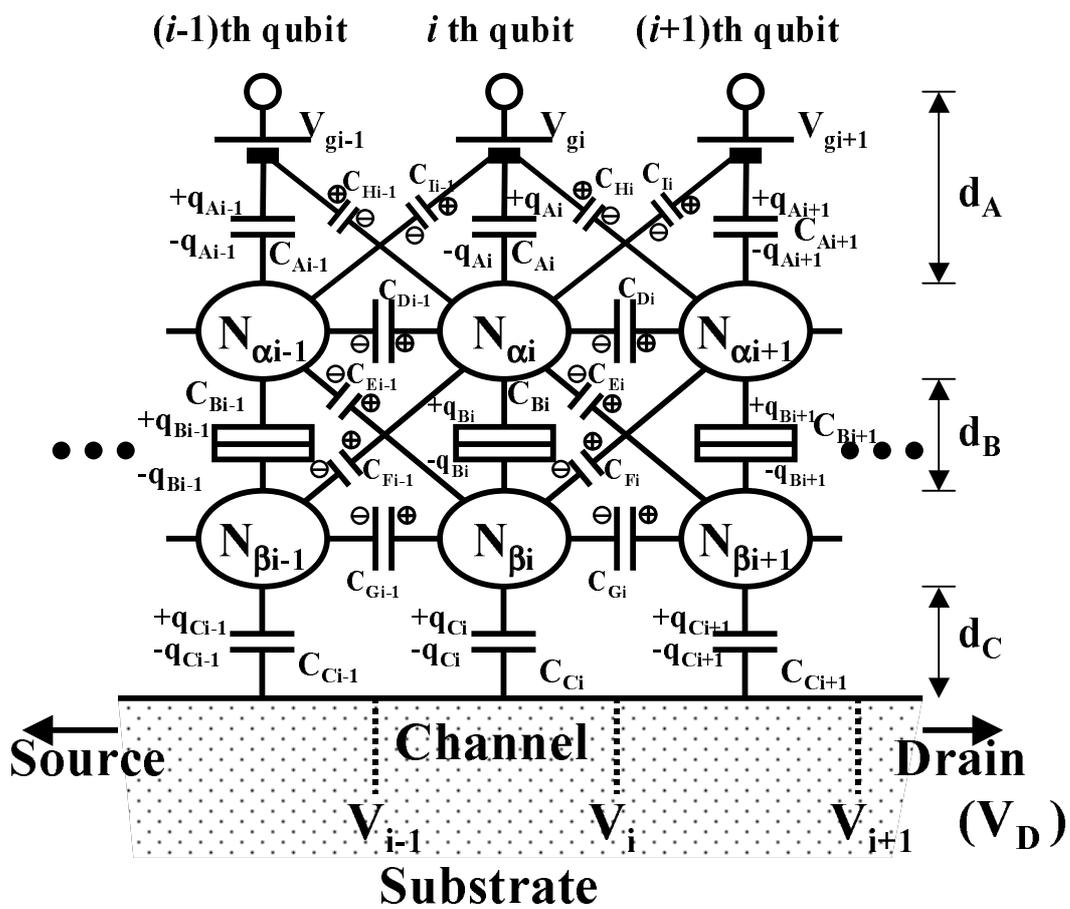
\begin{figure}
\caption{Schematic of the capacitance network of a one-dimensionally 
coupled quantum-dot array. The capacitances $C_{{\rm A}i}$, $C_{{\rm B}i}$, 
and $C_{{\rm C}i}$, the gate $V_{{\rm g}i}$ and the two quantum dots 
constitute the $i$th qubit.
The excess electron moves between the two quantum dots in a qubit 
via the tunneling barrier, and 
electron transfer between qubits is prohibited.}
\label{fig_cap}
\end{figure}

The fundamental idea of the quantum-dot qubits based on charged states is as follows. 
A qubit is composed of two quantum dots coupled 
via a thin tunneling barrier and  a gate electrode that is 
attached on a thick insulating material. 
This is a bistable well structure \cite{Landauer2}
where the electronic quantum state of the coupled dots is controlled
by the gate bias. %
First, we consider the one-dimensional arrayed qubits(Fig. \ref{fig_cap}).
The quantum dots are, {\it e.g.}, Si nanocrystals \cite{Tiwari,Guo,Ohba} 
or GaAs dots \cite{Zhitenev,Oosterkamp}. 
The quantum dots are assumed to be sufficiently small for  
charging effects to be observed. 
$N_{\alpha i}$ and $N_{\beta i}$ are the numbers of excess electrons 
from the neutral states in two quantum dots. 
One excess charge is assumed to be inserted 
from a substrate first and to stay in the two-coupled dots ($C_{Bi} > C_{Ci}$). 
When the excess charge exists in the upper dot 
and lower dot, we call them $|0\rangle=|\! \uparrow \rangle$ state 
and  $|1\rangle=|\! \downarrow \rangle$ state, 
respectively. 
The quantum logic gates are assumed to be operated in the region 
where the $|0 \rangle$ and $|1 \rangle$ states are near-degenerate, 
then the system becomes a two-state system\cite{Averin,Shnirman,Nakamura}. 
The qubits are arranged one-dimensionally and 
are capacitively coupled. 

Here we show the Hamiltonian of the one-dimensionally arrayed coupled 
quantum dots for quantum computation by starting from the tunneling Hamiltonian:
\begin{equation}
H=\sum_{i=1}^{N} (t \hat{a}_{i}^\dagger \hat{b}_{i} 
\!+\! t^* \hat{b}_{i}^\dagger \hat{a}_{i} 
\!+\! \epsilon_{\alpha i}  \hat{a}_{i}^\dagger \hat{a}_{i} 
\!+\! \epsilon_{\beta i}  \hat{b}_{i}^\dagger \hat{b}_{i} )
+ H_{\rm ch}, 
\end{equation}
where $\hat{a}_{i}$ ($\hat{b}_{i}$) describes the annihilation operator 
when the excess electron exists in the upper (lower) dot, and
$\epsilon_{\alpha i}$ ($\epsilon_{\beta i}$) shows the electronic energy 
of the upper (lower) dot. 
There is no restriction on the number of coupled quantum dots, $N$.
$H_{\rm ch}$ is the charging energy that includes the interaction between qubits. 
Because we consider the Coulomb blockade in the weak coupling region, 
the resistance of the interdot tunneling barrier should be larger than 
$R_K$($=h/e^2 \sim$ 25.8k$\Omega$) and the operational temperature be 
less than the charging energies. 
Thus, the operational speed should be less than the $CR$ constant 
of the capacitance network so that the double well potential 
profile generated by the charging energy is effective (adiabatic 
region). The criterion of the operation is discussed below in detail.
As for the interaction between qubits,  
the distribution of the excess charge is considered to be 
antiferromagnetic due to the repulsive Coulomb interaction. 
We can show that this interaction between qubits is 
an Ising interaction by minimizing the 
general formulation of the charging energy:
\begin{eqnarray}
\lefteqn{H_{\rm ch}\!=\! \! 
\sum_{i=1}^{N}\! \left\{ \frac{q_{{\rm A}i}^2}{2C_{{\rm A}i}}
\!+\!\frac{q_{{\rm B}i}^2}{2C_{{\rm B}i}}\!
+\!\frac{q_{{\rm C}i}^2}{2C_{{\rm C}i}}\!-\!q_{{\rm A}i}V_{{\rm g}i} \right\} }
\nonumber \\
\!&+&\!\sum_{i=1}^{N-1}\! \left\{\frac{q_{{\rm D}i}^2}{2C_{{\rm D}i}}\!
+\!\frac{q_{{\rm E}i}^2}{2C_{{\rm E}i}}
\!+\!\frac{q_{{\rm F}i}^2}{2C_{{\rm F}i}}\!
+\!\frac{q_{{\rm G}i}^2}{2C_{{\rm G}i}}\right\},
\nonumber \\
\!&+&\!\sum_{i=2}^{N}\! \left\{\frac{q_{{\rm H}i}^2}{2C_{{\rm H}i}}
\!-\!q_{{\rm H}i}V_{{\rm g}i-1} \! \right\}
\!+\!\sum_{i=1}^{N-1}\! \left\{ \frac{q_{{\rm I}i}^2}{2C_{{\rm I}i}}
\!-\!q_{{\rm I}i} V_{{\rm g}i+1}\! \right\},
\label{eqn:H_ch}
\end{eqnarray}
with constraints 
$-N_{\alpha i}\!=\!q_{{\rm A}i}\!-\!q_{{\rm B}i}\!+\!q_{{\rm D}i}
\!-\!q_{{\rm D}i-1}\!+\!q_{{\rm E}i}\!-\!q_{{\rm F}i-1}\!+\!q_{{\rm H}i-1}
\!+\!q_{{\rm I}i}$ and  
$-N_{\beta  i}\!=\!q_{{\rm B}i}\!-\!q_{{\rm C}i}\!+\!q_{{\rm G}i}
\!-\!q_{{\rm G}i-1}\!-\!q_{{\rm E}i-1}\!+\!q_{{\rm F}i}$ ($i=1,..,N$).
The last line of Eq.(\ref{eqn:H_ch}) shows the effects of 
other gate electrodes on the $i$th qubit (cross-talk).
By using Lagrange multiplier constants, 
the energy of the system is obtained as a function 
of the relative excess charge $n_i \equiv N_{\alpha_i}-N_{\beta_i}$
and the total excess charge of the two quantum dots 
$N_i \equiv  N_{\alpha_i}+N_{\beta_i}$(=1 ):
\begin{eqnarray}
\lefteqn{
H_{\rm ch} \cong \! \! \sum_{i=1}^N \frac{C_{bi}}{2D_i} \! \left( \!
n_{i}\!+\! \frac{C_{ci}}{C_{bi}} N_i \!
+\! \left( 1 \!+\! \frac{C_{ci}}{C_{bi}} \right) 
Q_V \right)^2  } \nonumber \\
&+&\sum_{i=2}^{N}\!\frac{2}{D_i D_{i-1}}
(C_{bi}C_{bi-1}C_{ei-1}\!+\!C_{ci}C_{ci-1}C_{di-1})n_{i}n_{i-1}, 
\label{eqn:u2}
\end{eqnarray}
with 
$C_{ai}$ $\equiv$ $C_{Ai}\!+C_{Ci}\!+4C_{Bi}$
$\!+\!2(C_{Di}\!+C_{Ei})\!+C_{Hi}\!+C_{Ii}$, 
$C_{bi}$ $\equiv$ $C_{Ai}\!+C_{Ci}$$+\!2(C_{Di}\!+C_{Ei})$
$+\!2(C_{Di-1}\!+C_{Ei-1})\!+C_{Hi}\!+C_{Ii}$, 
$C_{ci}$ $\equiv$ $C_{Ci}\!-C_{Ai}$$+C_{Hi}\!+C_{Ii}$, 
$C_{di}$ $\equiv$ $C_{Di}\!+\!C_{Ei}$, 
$C_{ei}\equiv C_{Di}-C_{Ei}$ ($C_{Di}\!=\!C_{Gi}$, $C_{Ei}\!=\!C_{Fi}$) and 
$D_i \equiv C_{ai}C_{bi}-C_{ci}^2$
 and $Q_V \equiv C_{Ai}V_{{\rm g}i} \!+\! 
C_{Hi-1}V_{{\rm g}i\!-\!1} \!+\! C_{Ii}V_{{\rm g}i+1}$.  
It is assumed that the coupling between qubits is smaller than 
that within a qubit and we neglect higher order terms than 
$(C_{di}^2/D_i)^2 (\ll 1)$. 
This assumption is valid when the distance between the quantum dots 
in different qubits is larger than that between quantum dots in a qubit. 
From Eq.(\ref{eqn:u2}), we define the characteristic charging energy 
of the system as $E_C \equiv C_{b}/(2D)$. 
We consider the gate voltage region, where the $n_{i}$=-1 state and 
$n_{i}$=1 state are near-degenerate as in Ref.\cite{Averin,Shnirman,Nakamura}, 
and we obtain the Hamiltonian of the coupled quantum dot system:
\begin{equation}
H =\sum_{i=1}^{N} [t \hat{a}_{i}^\dagger \hat{b}_{i} 
+t^* \hat{b}_{i}^\dagger \hat{a}_{i} +
\Omega_i \hat{I}_{iz}] \!+\! \sum_{i=1}^{N-1} 
J_{i,i+1} \hat{I}_{iz} \hat{I}_{i+1z}, 
\label{eqn:H_sys}
\end{equation}
where $|\!\! \uparrow_i \rangle \! \! =\! \! |n_i\!=\!1 \rangle$ and 
$|\! \downarrow_i \rangle \! =\! |n_i\!=\!-1 \rangle$, 
$\hat{I}_{iz} = (\hat{a}_{i}^\dagger \hat{a}_{i}\!-\! 
\hat{b}_{i}^\dagger \hat{b}_{i} )/2$, and
\begin{eqnarray}
\Omega_i &=&\frac{4C_{Ci}}{D_i} \!
\left[ Q_V \! -\! Q_V^{\rm res} \right] , \\
J_{i,i+1}&=&\frac{1}{2D_i D_{i\!+\!1}}
[C_{bi}C_{bi+\!1}C_{ei+\!1}\!+\!C_{ci}C_{ci+\!1}C_{di+\!1}].
\end{eqnarray}
$Q_V^{\rm res}$ includes $\epsilon_{\alpha i}$ and $\epsilon_{\beta i}$ 
and shows the gate voltage when the $|0 \rangle$ and $|1 \rangle$  degenerate 
(on resonance). 
We control the time-dependent quantum states of the qubits in the vicinity of 
on resonant gate bias by applying a gate voltage such 
as $V_i (\tau) =V_i^{\rm res}+ v_i(\tau)$ where $V_i^{\rm res}$ is the 
gate bias of on resonance.   
In this on resonant region, a transformation of the coordinate:
\begin{equation}
\left(
\begin{array}{c}
\hat{\alpha}_{+i} \\
\hat{\alpha}_{-i} 
\end{array}
\right)
\equiv U_{0} 
\left(
\begin{array}{c}
\hat{a}_{i} \\
\hat{b}_{i} 
\end{array}
\right)
, \ \ 
U_{0}\equiv \frac{1}{\sqrt{2}}
\left(\begin{array}{cc}
1 & 1\\
1 & -1  
\end{array}
\right).
\end{equation}
is convenient (we neglect the phase of $t_i$ for simplicity). 
Then the Hamiltonian Eq.(\ref{eqn:H_sys}) can be described as  
\begin{equation}
H(\tau) \!=\! \sum_{i=1}^N [
2t_{i}\hat{I}'_{zi} \!-\!  \Delta_i (\tau)
\hat{I}'_{xi}] 
\!+\! \sum_{i=1}^{N-1} 
J_{i,i+1} \hat{I}'_{xi}\hat{I}'_{xi+1}
\label{eqn:H_on}
\end{equation}
where 
\begin{eqnarray}
I'_{xi} &\!\equiv\!& (\alpha_{+i}^\dagger \alpha_{-i} \!
+ \! \alpha_{-i}^\dagger \alpha_{+i})/2, 
I'_{yi} \!\equiv\! i(\alpha_{+i}^\dagger \alpha_{-i} \!
- \! \alpha_{-i}^\dagger \alpha_{+i})/2, \nonumber \\
I'_{zi} \! &\equiv& \!(\alpha_{+i}^\dagger \alpha_{+i}
\!-\! \alpha_{-i}^\dagger \alpha_{-i})/2,  
\end{eqnarray}
and
\begin{equation}
\Delta_i (\tau) = \frac{4C_{Ai}C_{Ci}}{D_i} 
[v_{i}(\tau)  + \delta v_{i,{\rm cr}}(\tau)]. 
\end{equation}
$\delta v_{i{\rm cr}} (\tau) \equiv 
[C_{Hi-1}v_{i-1}(\tau)\!+\! C_{Ii}v_{i+1}(\tau)]/C_{Ai}$
is a cross-talk term and is an effect of other gate electrodes. 
Hamiltonian (\ref{eqn:H_on}) is an NMR Hamiltonian 
when we regard $\omega_i =2 t_i$ as a Zeeman energy 
and $\Delta_i (\tau)$ as a transverse magnetic field, 
if $\omega > \Delta_0 \gg J$. 
Thus, when $\Delta_i (\tau)=\Delta_{0i} \cos (\omega_i \tau + \delta_i)$ and 
we take a rotating wave approximation by the unitary transformation 
$U_{\rm rwa}(\tau) =\exp (-i \sum_{i=1}^{N} \omega_i \tau I_{iz}' )$, 
we have 
\begin{eqnarray}
\lefteqn{ H_{\rm rwa}\!\!\!=\! -\sum_{i=1}^N 
\! \frac{\Delta_{i0}}{2} [\hat{I}'_{xi} \cos \delta_i 
+\hat{I}'_{yi} \sin \delta_i ]  }\nonumber \\
\!&+&\! \sum_{i=1}^{N-1} \frac{J_{i,i+1}}{2} 
[\hat{I}'_{xi}\hat{I}'_{xi+1}+\hat{I}'_{yi}\hat{I}'_{yi+1}].
\label{eqn:H_rwa}
\end{eqnarray}
The pulse process is carried out when the oscillating electric field is applied 
($\Delta_0 \neq 0$) where the interaction term is considered to be 
able to be neglected because $\Delta_0 \gg J$.  
For example, $\pi_z$-pulse in the $(\hat{a}_{i},\hat{b}_{i})$ basis 
corresponds to $\pi_x$-pulse representation in this 
$(\hat{\alpha}_{+i},\hat{\alpha}_{-i})$ basis 
and is carried out when $\delta=0$ and $\Delta_0 \tau/2 =\pi$.  
In general, rotation 
$R_{\gamma} (\theta) \equiv e^{\ i\theta I_\gamma}$ $(\gamma=x,y,z)$ 
in the $(\hat{a}_{i},\hat{b}_{i})$ basis is interpreted 
by $R_{i}' (\theta) =U_0^{-1} R_{i} (\theta) U_0$ 
in the $(\hat{\alpha}_{+i},\hat{\alpha}_{-i})$ basis. 
The evolution process is carried out when the oscillating voltage is 
not applied ($\Delta_0$ =0). 
A two-qubit CNOT gate is described by  
a pulse sequence of the form $\hat{U}_{\rm CNOT}^{ij} \propto$
$U_0^{-1}R_{ix}'(\pi/2) R_{jy}'(\pi/2) R_{jx}'(\pi/2) R_{ij}(-\pi) R_{jy}'(-\pi/2)U_0$
\cite{Linden,Vandersypen}. 
The 2-body interaction $R_{ij}(-\pi)=e^{-\pi \hat{I}_{iz}\hat{I}_{i+1z}}$ 
is obtained by using an average Hamiltonian theory\cite{Ernst}.
Here we apply the ``Carr-Purcell" sequence $\tau-\pi_y-2\tau-\pi_y-\tau$
to the Hamiltonian Eq.(\ref{eqn:H_on}) in order to average the effect of Zeeman term to zero. 
Thus, we can show that the Hamiltonian of 
the coupled quantum dot system can be reduced to that of 
the weakly coupled system in NMR and quantum operations 
can be carried out in a manner similar to that of operations in 
NMR quantum computers.

Similarly,  
we can show that the Hamiltonian of the two-dimensionally arrayed coupled dots 
is reduced to that of an artificial Ising system controllable by electric fields.
Here we consider the case where there are only nearest-neighbor 
capacitive couplings ($C_E$ and $C_F$ are neglected in the above formulation). 
If we express the coupling strength between the sites ${\bf i}\equiv (i,j)$ and 
${\bf i\!+\!x}\equiv (i\!+\!1,j)$ as $J_{\bf i}^x$ and
that between the sites ${\bf i}$ and ${\bf i\!+\!y}\equiv (i,j+1)$ as $J_{\bf i}^y$ 
($1\le i \le N_x$, $1\le j \le N_y$: $N=N_xN_y$), 
we have a Hamiltonian similar to Eq. (\ref{eqn:H_sys}) where 
\begin{eqnarray}
J_{\bf i}^x&=&\frac{2}{D_{\bf i}D_{\bf i+x}}(C_{b{\bf i}}C_{b{\bf i+x}}
+C_{c{\bf i}}C_{c{\bf i+x}})C_{D{\bf i}}^x, \\
J_{\bf i}^y&=&\frac{2}{D_{\bf i}D_{\bf i+y}}(C_{b{\bf i}}C_{b{\bf i+y}}
+C_{c{\bf i}}C_{c{\bf i+y}})C_{D{\bf i}}^y,  
\end{eqnarray}
with 
$D_{\bf i} \equiv C_{a{\bf i}}C_{b{\bf i}}-C_{c{\bf i}}^2$, 
$C_{a{\bf i}}$ $\equiv$ $C_{A{\bf i}}\!+\!C_{C{\bf i}}\!+\!4C_{B{\bf i}}$, 
$C_{b{\bf i}}$ $\equiv$ $C_{A{\bf i}}\!+\!C_{C{\bf i}}$, 
$C_{c{\bf i}}$ $\equiv$ $C_{C{\bf i}}\!-\!C_{A{\bf i}}$.  
$C_{D{\bf i}}^x(=C_{G{\bf i}}^x)$ and $C_{D{\bf i}}^y(=C_{G{\bf i}}^y)$ 
are capacitances between qubits in $x$-direction 
and $y$-direction, respectively. 

Here we summarize the criterion for realizing the above-mentioned scheme. 
First of all, the Coulomb blockade should be effective 
at the operational temperature 
$T$ and we have $T \ll E_C$.  
In view of time-dependent operation, the excess charge 
is assumed to be affected by the electronic potential 
generated by the capacitance network. 
Therefore, all quantities concerning time evolution should be smaller 
than the $CR$ constant of the network. Here we take $CR=C_{\rm int}R_{\rm int}$
where $C_{\rm int}(\equiv D/C_b)$ and $R_{\rm int}$ are capacitance and resistance 
of the interdot tunneling barrier in a qubit, respectively.
By including the condition for using the pulse sequence as mentioned above,  
we have a condition for the operation:
\begin{equation}
T \ll J \ll \Delta_0 < t  \ll (CR)^{-1}. 
\label{eqn:cond}
\end{equation}
We can roughly estimate this criterion by taking typical values, 
$r_0$ =2.5 nm (radius of a quantum dot),  
$d_{A}$$=$8 nm, $d_{B}$= 1.5 nm, $d_{C}$= 2.5 nm, and the distance 
between qubits $d_D$ is 12 nm ($\epsilon_{\rm ox}$ =4 (SiO${}_2$) 
and $\epsilon_{\rm Si}$=12), reflecting several experimental data 
\cite{Tiwari,Guo,Ohba}.
Using relations, 
$C_{A,C}$=$2\pi\epsilon_{\rm ox} r_0^2 /
(d_{A,C} \!+\!(\epsilon_{\rm ox}/\epsilon_{\rm Si})r_0)$ and 
$C_{B}$ $=$ $2\pi\epsilon_{\rm ox} r_0^2 /
(d_B\!$ $+\!2(\epsilon_{\rm ox}/\epsilon_{\rm Si})r_0)$, 
we obtain $E_C \sim 13$meV(150K), 
$t \sim 0.4$meV and $J\!\sim \! 0.1$ meV. 
If $R_{\rm int}$ is of the order of M$\Omega$, 
we obtain $(C_{\rm int} R_{\rm int})^{-1} \sim$ 3.1THz 
and greater than $t \sim$ 100GHz.  
Thus, the condition (\ref{eqn:cond}) is satisfied 
if the operational temperature should be much less than 1K.
If we could prepare $r_0=0.5$nm, $d_B=1.2$nm and $d_D=2$nm, 
we obtain $t \sim $120 K and $J \sim $ 90K and quantum calculations 
can be expected to be carried out at around liquid nitrogen temperature. 
The effects of cross-talk are of the order of 
$C_{Hi}/C_{Ai} \sim d_{Ai}/d_{Di}$ and cannot be neglected even when
gate electrodes are set closer to the corresponding quantum dots. 
However, we can control the cross-talk effects 
by adjusting $v_{i}(\tau)$ to obtain the required $\Delta_i (\tau)$.

At the near-degeneracy point, the system becomes a two-state system and 
the estimation of the decoherence (in order of $\mu$sec) discussed in 
Ref.\cite{Tanamoto} may be applicable. Until now, we have not known of 
any corresponding experimental 
data for the decoherence time\cite{Fujisawa}. The point is that 
various methods developed in NMR spectroscopy, 
such as the composite pulse method\cite{Ernst}, 
can be utilized to reduce the imperfections of the pulse 
and coherence transfer, 
that is, errors that are brought about in the 
operations. 
In addition, if the speed of quantum computations can be increased 
to much more than the shortest decoherence time, 
$\tau_c \!\sim\! \omega_c^{-1}$$ (\!\sim\! 10^{-14}$ s), 
group-theoretic approaches \cite{Viola2,Zanardi} 
for decreasing the decoherence will be effective. 
In this case, we have to reduce the corresponding $CR$ of the junction 
such that Eq.(\ref{eqn:cond}) holds 
and the cycle time of operations 
should be much less than $\tau_c$. 
Thus, the errors and the effects of decoherence of a coupled quantum-dot system 
will be reduced by developing many contrivances.

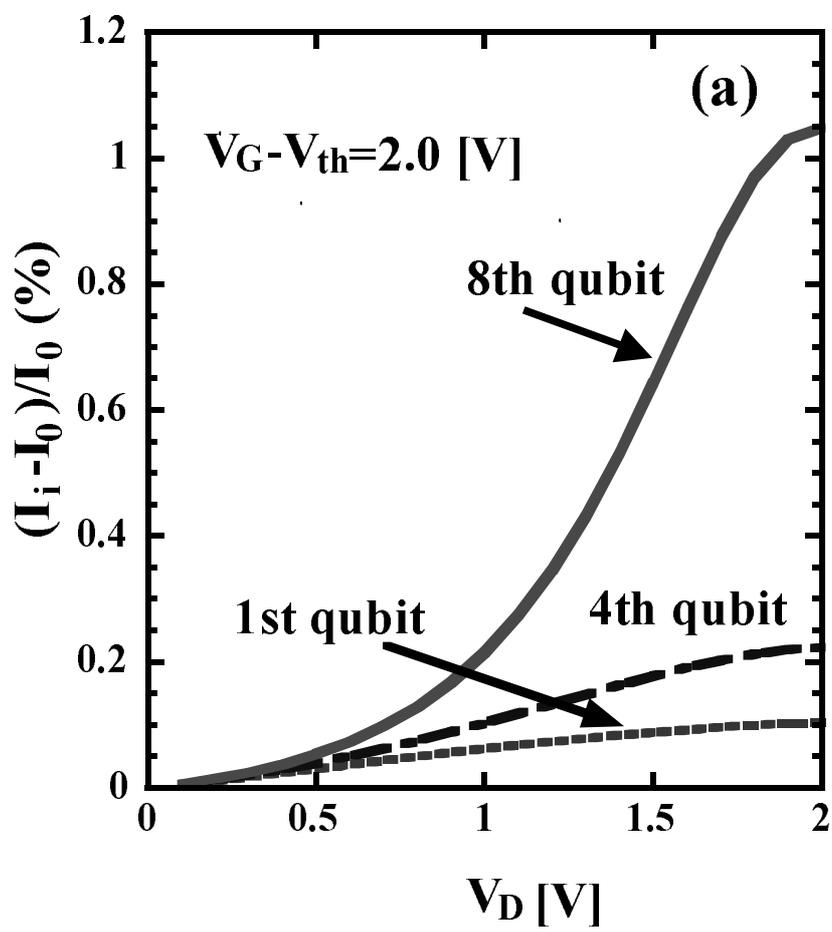
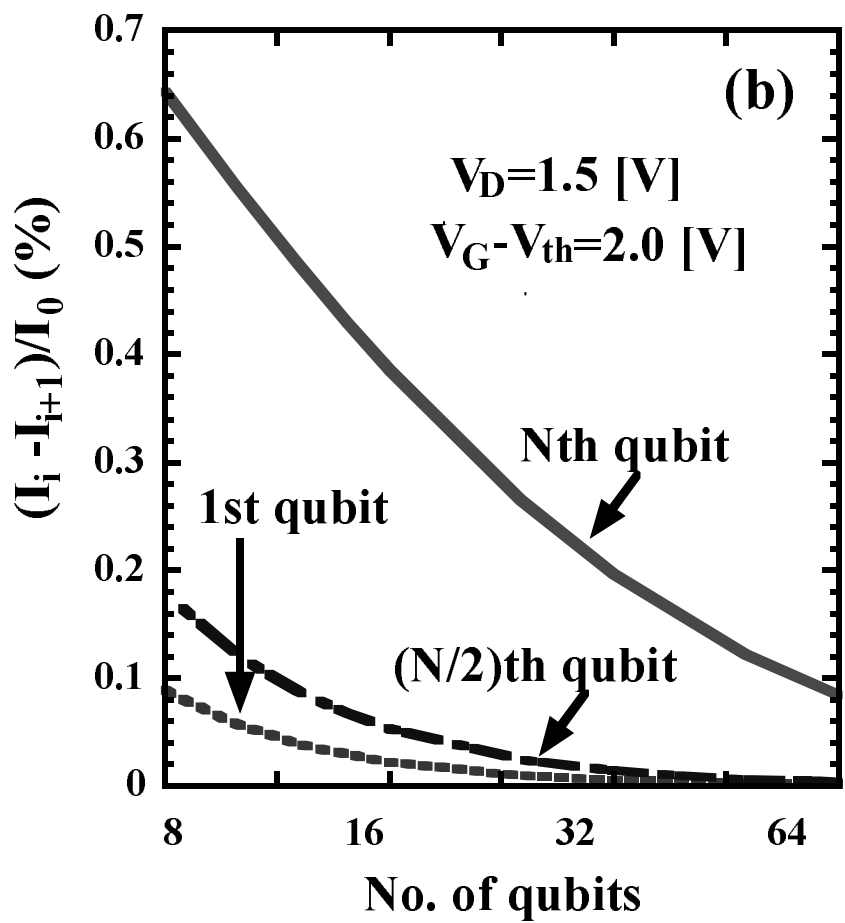
\begin{figure}
\caption{
$I_i$ represents the current where 
only the $i$th qubit shifts its threshold voltage 
$V_{{\rm th}}(0)\!+\!\Delta V_{{\rm th}i}$. 
$I_0$ represents the current where all qubits have the same 
threshold voltage $V_{{\rm th}}(0)$. 
(a) The ratio of current change $|I_i-I_0|/I_0$  as a function of $V_D$ in 
an 8-qubit quantum computer ($i\!=\!1$, $i\!=\!4$, and $i\!=\!8$). 
(b) The ratio of change $|I_i-I_{i+1}|/I_0$ as a function of the number of qubits,  
in the case $i=1$(near source), $i=N/2$ (middle) and $i=N-1$(near drain) 
at $V_D=1.5$V.
The efficiency  of detecting the quantum state of qubits is highest 
when the qubits near the drain change their states.
In (a) and (b), $V_{\rm g}-V_{\rm th}$=2V and  $\Theta$=0.3 V${}^{-1}$
in Eq.(7) in text.
The threshold shift is 10\% of $V_{\rm g}-V_{\rm th}$.}
\label{fig_I}
\end{figure}

Hereafter we consider the one-dimensionally arrayed qubits for simplicity. 
Next, we quantitatively illustrate the {\it reading out} process 
based on an FET structure (Fig.\ref{fig_cap}).  
Measurement is carried out, after quantum calculations, by 
applying a finite bias $V_{\rm D}$ between the source and drain\cite{Tanamoto}. 
The detection mechanism is such that the change of the charge distribution 
in a qubit induces a threshold voltage shift $\Delta V_{{\rm th}}$
of the gate voltage above which the channel current flows in the substrate. 
$\Delta V_{{\rm th}}$ is of the order of $e d_q/\epsilon_{\rm ox}$($d_q$ is a distance 
between the centers of two quantum dots in a qubit)\cite{Guo}. 
The effect of $\Delta V_{{\rm th}}$ 
differs depending on the position of the qubit, 
because the width of the depletion region in the substrate 
changes gradually from source to drain. 
A simplified model of a metal-oxide semiconductor 
field-effect-transistor (MOSFET) including 
velocity saturation effects 
($\Theta$ in Eq. (\ref{eqn:MOS}))\cite{Hoefflinger} 
is used for the current that flows under the $i$th qubit
\begin{equation}
I_{\rm D}^{(i)}\!=\! \Lambda \frac{
[V_{{\rm g}i}\!-\!V_{{\rm th}i}](V_i \!- \!V_{i-1})
\!-\!(1/2)\eta_i (V_i^2\!-\!V_{i-1}^2)}
{ 1\!+\!\Theta(V_i \!-\! V_{i-1})},
\label{eqn:MOS}
\end{equation}
where $\Lambda  \equiv Z \mu_0 C_0/L_0$ ($Z$ is the channel width, 
$\mu_0$ is the mobility, $L_0$ is the channel length of one qubit, 
and $C_0$ is the capacitance of the gate insulator),  
$\eta_i \equiv 1+\zeta_i$
where $\zeta_i$ is determined by the charge of the surface depletion region, 
and $V_i$ is the voltage of $i$th qubit to be determined 
from $V_{\rm N}\!=\!V_{\rm D}$ and 
$I_{\rm D}^{(1)}\!=\!I_{\rm D}^{(2)}\!=\! \cdots\!=\!I_{\rm D}^{(N)}$. 
The detection should be carried out before the change of 
the quantum states of the qubits. For this purpose, $\Lambda$ should be 
as large as possible\cite{Tanamoto}. 
The threshold voltage of $i$th qubit $V_{{\rm th}i}$ is given by 
$V_{{\rm th}i} \!=\! V_{\rm th}(\zeta_i)\!+\! \Delta V_{{\rm th}i}$.  
Figure \ref{fig_I}(a) shows the ratio of the current change 
$|I_i$-$I_0|$/$I_0$ as a function of $V_D$ in 8 qubits  
with $\zeta_i=0$, 
where $I_0$ is the initial current and $I_i$ is the current when 
$V_{{\rm g}i}$-$V_{{\rm th}i}$ of $i$th qubit changes by 10\%. 
This ratio is largest 
for a qubit near the drain ($i\!=\!8$) 
because it has the narrowest inversion layer. 
To show how to distinguish qubits, 
we compare the current where only the $i$th qubit 
shifts its threshold voltage with that where only the ($i$+1)th qubit 
shifts its threshold voltage. 
Figure \ref{fig_I}(b) shows the results for (i)$i$=1, (ii)$i$=$N/2$, and 
(iii)$i$=$N-1$, with the same voltage shift 
as in Fig.\ref{fig_I}(a). 
The maximum allowable number of arrayed qubits depends on 
the sensitivity of the external circuit to the channel current. 
When the density of acceptor in the substrate is of the order of 
10${}^{17}$ cm${}^{-3}$, the number of acceptors below one qubit 
is less than one. If this effect is represented by $\zeta_i$ 
as a random number, the above ratios may increase or decrease 
and the overall features are similar to those in Fig.\ref{fig_I}. 
To construct a large qubit array, additional dummy qubits 
are needed over the source and drain so that separated qubits 
on different FETs are connected.

In conclusion, we have theoretically shown that the Hamiltonian for a weakly 
coupled quantum-dot array in the Coulomb blockade regime 
is reduced to that for NMR spectroscopy. 
The flexible quantum information processing developed in the 
NMR quantum computer and a variety of error-correction 
methods in conventional NMR are 
applicable to the quantum-dot system that is considered to be the 
most feasible system in view of the present technology.  
The difficulty of scalability in the NMR computer, 
namely that the overlap of pulses 
restricts the number of qubits\cite{Bennett}, is overcome in the 
operations by individual gate electrodes.  
The disadvantage of the short decoherence time is also compensated for
by another advantage, namely that the measurement procedure is compatible 
with classical circuits. 
It is expected that a transmitting loss of signals to classical circuits 
will decrease due to this compatibility.
Recently, nanofabrication of two-dimensionally distributed 
self-aligned Si doubly-stacked dots has been successfully realized 
in the form of a non-volatile memory device\cite{Ohba} and the detailed 
analysis of the behavior of electrons, such as an artificial antiferromagnet, 
is expected to be performed. 
The $N$-qubit system composed of arrayed quantum dots is expected to 
be developed as a result of further advances in fabrication technologies.  

The author thanks N. Gemma, R. Katoh, K. Ichimura, J. Koga, R. Ohba and 
M. Ueda for useful discussion.


\normalsize

\end{document}